\documentclass[preprint,aps,draftams,math,amssymb,superscriptaddress,unsortedaddress]{revtex4-1}
\usepackage{graphicx}
\usepackage{dcolumn}
\usepackage{bm}
\usepackage[usenames,dvips]{color}
\usepackage{gensymb}

\begin{document}

\preprint{APS/123-QED}

\title{Theory of the ferromagnetism in Ti$_{1-x}$Cr$_{x}$N solid solutions}

\author{B. Alling}
\email{bjoal@ifm.liu.se}
\affiliation{Department of Physics, Chemistry and Biology (IFM), \\
Link\"oping University, SE-581 83 Link\"oping, Sweden.}
\date{\today}

\begin{abstract}
First-principles calculations are used to investigate the magnetic properties of Ti$_{1-x}$Cr$_{x}$N solid solutions. We show that the magnetic interactions between Cr spins that favour antiferromagnetism in CrN is changed upon alloying with TiN leading to the appearance of ferromagnetism in the system at approximately $x \leq 0.50$ in agreement with experimental reports. Furthermore we suggest that this effect originates in an electron density redistribution from Ti to Cr that decreases the polarization of Cr d-states with t$_{2g}$ symmetry while it increases the polarization of Cr d-states with e$_g$ symmetry, both changes working in favour of ferromagnetism.

\end{abstract}

\maketitle

\section{Introduction}
The transition metal nitrides TiN and CrN have become two of the main building blocks of designed multicomponent hard coatings, such as the solid solutions Ti$_{1-x}$Al$_{x}$N and Cr$_{1-x}$Al$_{x}$N, as well as nanocomposites like TiN/SiN$_{x}$, that have revolutionized the cutting tool and coatings industry over the last two decades~\cite{Mayrhofer2006Rev}. Mixing those two nitrides into the Ti$_{1-x}$Cr$_{x}$N solutions however, have been found not to further increase hardness~\cite{Hones1998} but is of interest for corrosion protection in e. g. fuel cell applications~\cite{Choi2009}. However, arguably the most intriguing aspect of this material from both a fundamental physics point of view as well as for possible future applications, is related to its surprising magnetic behaviour. 

CrN is well known to display antiferromagnetic ordering below a N\'eel temperature of about 280K~\cite{Corliss1960, Browne1970}. Associated with the magnetic ordering is a cubic (B1)  to orthorhombic structural distortion~\cite{Corliss1960}. TiN, also crystalizing in the B1 structure, on the other hand is a non-magnetic system showing superconductivity at low temperatures~\cite{Hardy1954}. Other 3d transition metal mono-nitrides like MnN and B1-FeN~\cite{Hinomura1997, Suzuki2000, Suzuki2001}, are antiferromagnetic.  Thus, the discovery by Aivazov and Gurov of \emph{ferromagnetism} in the solid solution system B1-Ti$_{1-x}$Cr$_{x}$N~\cite{Aivazov1975} is remarkable. This effect was re-discovered and investigated in depth almost thirty years later by Inumaru~\emph{et al.} in thin films~\cite{Inumaru2004, Inumaru2007} and in bulk samples~\cite{Inumaru2008} pointing out its possible importance for magnetoresistance applications. It is also easy to imagine the general usability of a ferromagnetic nitride material in spintronics, as it should be readily incorporable in nitride based semiconductor devices. The Curie temperature, showing a maximum of 140~\cite{Inumaru2007}-170~\cite{Aivazov1975}~K at about $x=0.50$, is however not high enough for most applications and further material development is needed to obtain the elusive hard nitride material that is ferromagnetic at room temperature. For such a purpose it is important to understand the underlaying physical mechanism that is responsible for the surprising appearance of ferromagnetism in Ti$_{1-x}$Cr$_{x}$N with $x \leq 0.5$, as well as its rather abrupt disappearance at $x > 0.5$ in Refs.~\cite{Inumaru2007, Inumaru2008}. 

Filippetti \emph{et. al} studied pure CrN using first-principles local spin density approximation methods and mapped the energies of a few different antiferromagnetic (AFM) and ferromagnetic (FM) configurations onto a two parameter Heisenberg Hamiltonian~\cite{Filippetti2000, Filippetti1999}. The nearest neighbor interaction was found to be negative favoring AFM while the second nearest neighbor was positive, although weaker in magnitude,  favoring FM. The overlap of half-filled non-bonding Cr 3d orbitals with t$_{2g}$ symmetry was suggested to give rise to the AFM coupling between moments on nearest neighbor Cr atom while a competition between superexchange and double exchange resulted in the net FM coupling between next-nearest neighbor Cr atoms~\cite{Filippetti2000}. Inamura~\emph{et al.} built on these discussions and suggested that their experimental findings of ferromagnetism in Ti$_{1-x}$Cr$_x$N could be due to a more pronounced weakening effect of Ti dilution on the Cr-Cr AFM nearest neighboring interaction as compared to the Cr-N-Cr FM interactions~\cite{Inumaru2007}. However, no quantitative explanation of this phenomena has yet been suggested. 

In this work we perform a thorough theoretical investigation, using first-principles calculations, of the magnetism in Ti$_{1-x}$Cr$_{x}$N substitutionally disordered solid solutions. We calculate the energies of different relevant magnetic states as a function of composition, derive the magnetic exchange interactions, simulate the phase diagram and suggest an explanation to the observed trends based on the electronic structure of the material.

\section{Calculational methodology}
We perform electronic structure calculations within two different but complementary density functional theory frameworks. Firstly the projector augmented wave (PAW) method~\cite{Blochl1994} as implemented in the Vienna \emph{ab-initio} simulation package (VASP)~\cite{Kresse1993, Kresse1999} is used. We use both the generalized gradient approximation (GGA)~\cite{Perdew1996}, and a combination of the local density approximation with a Hubbard Coulomb term (LDA+U)~\cite{Dudarev1998} to account for exchange correlation effects. The Hubbard term was applied to the Cr and Ti 3d-orbitals and the value of the effective U (U-J) was chosen to 3 eV in line with the findings of increased agreement with experimental band structure measurements as compared to LDA and GGA for pure CrN~\cite{AllingThesis, Herwadkar2009}. A cut-off energy of 400 eV is used in the plane wave expansion of the wave functions.

Secondly we use a Green's function technique~\cite{Skriver1991, Abrikosov1993, Ruban1999} utilizing the Koringa-Kohn-Rostocker~\cite{Korringa1947, Kohn1954} method and the atomic sphere approximation (KKR-ASA)~\cite{Andersen1975, Andersen1984} together with the GGA functional for exchange-correlation effects. A basis set of s, p, and d muffin-tin orbitals was used to expand the wave functions.

In the PAW calculations a special quasirandom structure (SQS) method~\cite{Zunger1990} is used to model the substitutional disorder of Ti and Cr atoms on the metal sublattice of 96 atoms supercells with the compositions x= 0.25, 0.50, and 0.75. In the KKR-ASA simulations on the other hand, we apply the coherent potential approximation (CPA)~\cite{Soven1967, Velicky1968, Kirkpatrick1970} to analytically model the solid solution on a fine concentration grid of $\Delta x= 0.05$. We complement the CPA with a model~\cite{Ruban2002_a, Ruban2002_b} for treating electrostatic interactions between the alloy components due to charge transfer, something that will be shown to be non-negligible. The screening constants needed are calculated using the locally self-consistent Green's-function method~\cite{Abrikosov1996} of a supercell with compositon $x=0.50$.

Lattice parameters are calculated independently for each magnetic structure with the PAW method which is more reliable in this matter as compared to the ASA treatment of the one-electron potentials. In the KKR-ASA simulations we use lattice spacings corresponding to cubic spline interpolation between the values calculated for x=0.00 0.25, 0.50, 0.75, and 1.00 with the PAW method.  The equilibrium lattice parameter in the cubic phase is in practice almost independent of the magnetic state and follows closely to the Vegards rule between the values for TiN (4.255 \AA~in GGA, 4.248 \AA~in LDA+U, 4.24 \AA~in the experiment of Ref.~\cite{Hardy1954}) and CrN (4.149 \AA~in GGA, 4.133 \AA~in LDA+U and 4.148 \AA~in the experiment in Ref.~\cite{Herle1997}) in line with the experimental finding in Ref.~\cite{Inumaru2008}.

The following magnetic structures have been considered in the electronic structure calculations: In the PAW calculations we consider ferromagnetic and single [001]-layer antiferromagnetic (AFM[001]$_1$) tetragonal ordered Cr spin configurations on the underlying B1 cubic lattice for the chemical compositions x= 0.25, 0.50, 0.75, and 1.00.
To get a reference enthalpy value for the paramagnetic phase of pure CrN we use the SQS structure designed for Ti$_{0.5}$Al$_{0.5}$N in Ref.~\cite{Alling2007} to model Cr$^\uparrow_{0.5}$Cr$^\downarrow_{0.5}$N, a procedure used recently to investigate the bulk modulus of cubic CrN~\cite{Alling2010natmat}. We also calculate the energy of the orthorhombic phase with a double layer [011] AFM ordering (AFM[011]$_2$) that is the equilibrium state of CrN at low temperatures~\cite{Corliss1960} for the composition x=0.50, 0.75, and 1.00. In contrast to the large orthorhombic distortion of the AFM[011]$_2$~structure which is considered in this work, the tetragonal distortion of the lattice in the AFM[001]$_1$ case is minimal in terms of energy and geometry. In pure AFM[001]$_1$ CrN, the energy gained by tetragonal relaxation of the cubic lattice is less than 0.3 meV/f.u and these effects are thus neglected in this work. Local lattice relaxations where performed for all SQS calculations with the exception of the calculations of pure CrN with disordered local moments. 

In the KKR-ASA approach we have considered the ferromagnetic, the AFM[001]$_1$ antiferromagnetic, and a disordered local moments (DLM) configuration treated within the CPA~\cite{Gyorffy1985}. All KKR-ASA calculations are done on ideal lattice points of the cubic B1 structure neglecting local lattice relaxations. The KKR-ASA Green's function approach allows us to perform a a straight forward derivation of magnetic exchange interaction parameters through the application of the magnetic force theorem~\cite{Liechtenstein1984}. We use the disordered local moments reference state when, for each composition independently, deriving the exchange interactions $J_{ij}$ between Cr moments of the Heisenberg hamiltonian

\begin{equation}
H=-\sum_{i\neq j}J_{ij} \mathbf{e}_i\mathbf{e}_j,
\end{equation}

\noindent where $\mathbf{e}_{i}$ and $\mathbf{e}_{j}$ are unit vectors in the direction of the magnetic moment on site $i$ and $j$ respectively. This hamiltonian is then used to simulate the magnetic critical temperatures using a Heisenberg Monte Carlo simulation scheme capable of handling chemical disorder between magnetic and non-magnetic atoms on a lattice. The critical temperature was taken from the peak of the magnetic part of the specific heat. Possible magnetic interactions with small induced Ti and N moments are neglected in this procedure as is the effect of longitudinal spin fluctuations. The exchange interactions up to the 8:th coordination shell is included in the simulations. A convergence test for the two compositions x=0.50 and 0.30 showed that the inclusion of all interactions up to the 40:th coordination shell did not change the calculated critical temperature by more than 10 K, a value also representative for the convergence with respect to size of the simulation box and the number of sampling steps per temperature.

\section{Results and discussion}
\subsection{Energies of magnetic structures}
\begin{figure}
\includegraphics[angle=-90,width=0.92\columnwidth]{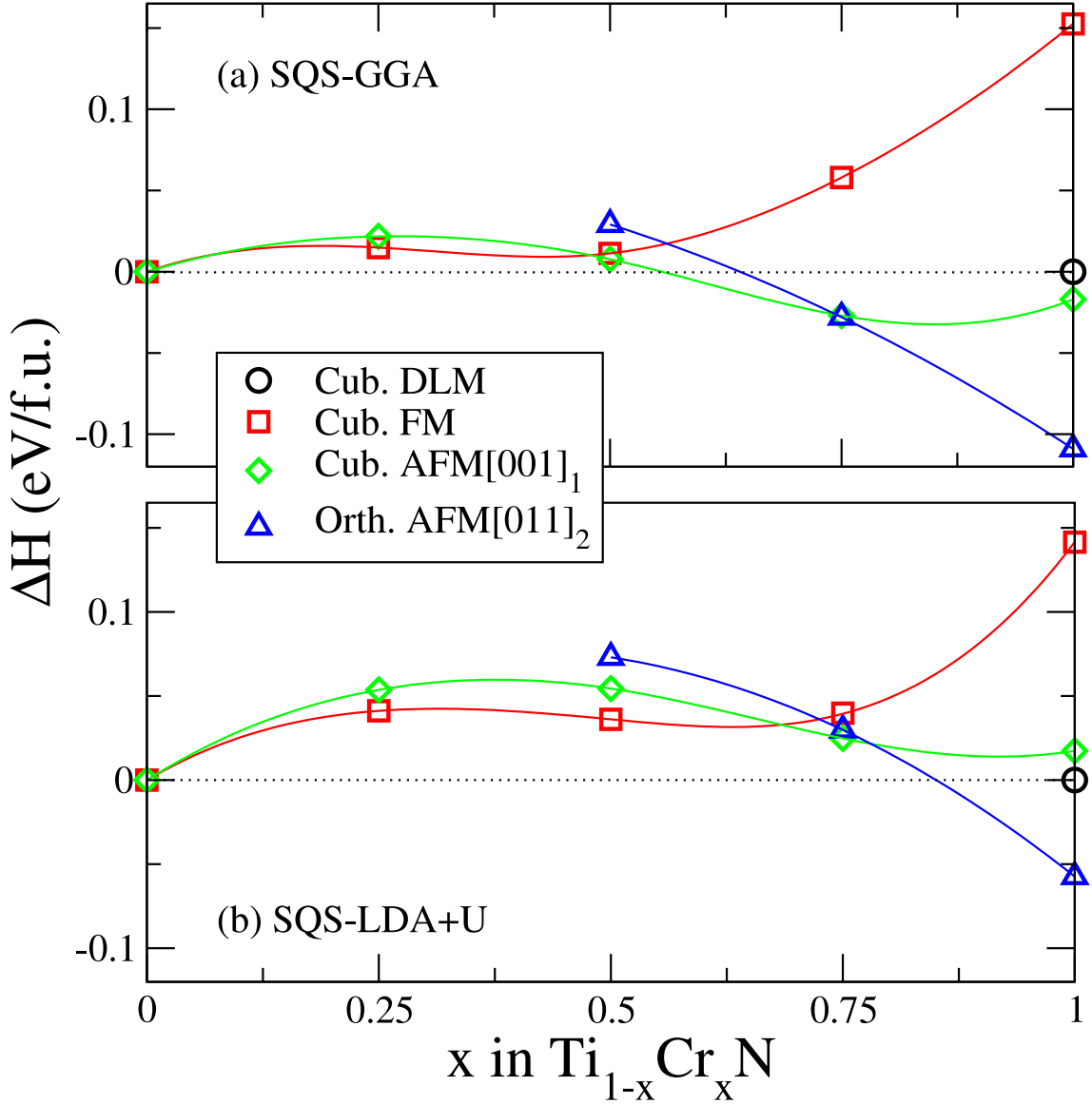}
\caption{\label{fig:SQS_H} (Color online) The calculated mixing enthalpies per formula unit (f.u.) of cubic ferromagnetic (squares), cubic antiferromagnetic [001]$_1$ (diamonds) and cubic disordered magnetic (circles), as well as orthorhombic antiferromagnetic [011]$_2$ (triangles) Ti$_{1-x}$Cr$_{x}$N solid solutions as calculated using the (a) PAW-SQS-GGA, and (b) PAW-SQS-LDA+U (U=3 eV) methods. The lines are guides to the eye.}
\end{figure}

Our first task is to investigate the relative energies of different magnetic structures of the Ti$_{1-x}$Cr$_{x}$N solid solutions as a function of composition. We do so by comparing the mixing enthalpies at zero pressure of the systems with respect to TiN and the cubic disordered magnetic state of CrN which is the experimentally found ground state for $x=1.00$ at room temperature. In Fig.~\ref{fig:SQS_H} we show the results of supercell calculations using the PAW-SQS method. We include three magnetic structures in this figure: the cubic FM, the cubic AFM[001]$_1$ and the orthorhombic antiferromagnetic [011]$_2$ structure. We employ both the GGA functional, previously used to study e. g. the Cr$_{1-x}$Al$_{x}$N system~\cite{Alling2007JAP} and the LDA+U with effective U=3 eV, shown to give increased agreement with experimental electronic structure of pure CrN~\cite{AllingThesis, Herwadkar2009}. Qualitatively the two approximations are in good agreement. In pure CrN the experimentally found low temperature ground state, orthorhombic AFM[011]$_2$, has the lowest energy while the cubic FM state is high in energy. As TiN is added to the system (decreasing $x$) the orthorhombic AFM[011]$_2$ state becomes less favorable while the relative energy of the FM state decrease. At about $x=0.75$ the cubic AFM[001]$_1$ magnetic state becomes lower in energy as compared to the orthorhombic state. This can be compared to the experimental work by Aivazov and Gurov who could identify an orthorhombic distortion in Ti$_{0.11}$Cr$_{0.90}$N, but not in Ti$_{0.20}$Cr$_{0.82}$N~\cite{Aivazov1975} (the slightly non-stoichiometric compositions are as stated in the reference).

At the composition $x=0.50$ the GGA-based calculations give all considered magnetic states similar energies with the cubic AFM[001]$_1$ being just below the cubic FM state. In the LDA+U calculations on the other hand the cubic FM state is clearly below the cubic AFM[001]$_1$ state at this composition. At even higher TiN content, represented by the composition $x=0.25$, the ferromagnetic state is lowest in energy also in the GGA calculations. 

\begin{figure}
\includegraphics[angle=-90,width=0.92\columnwidth]{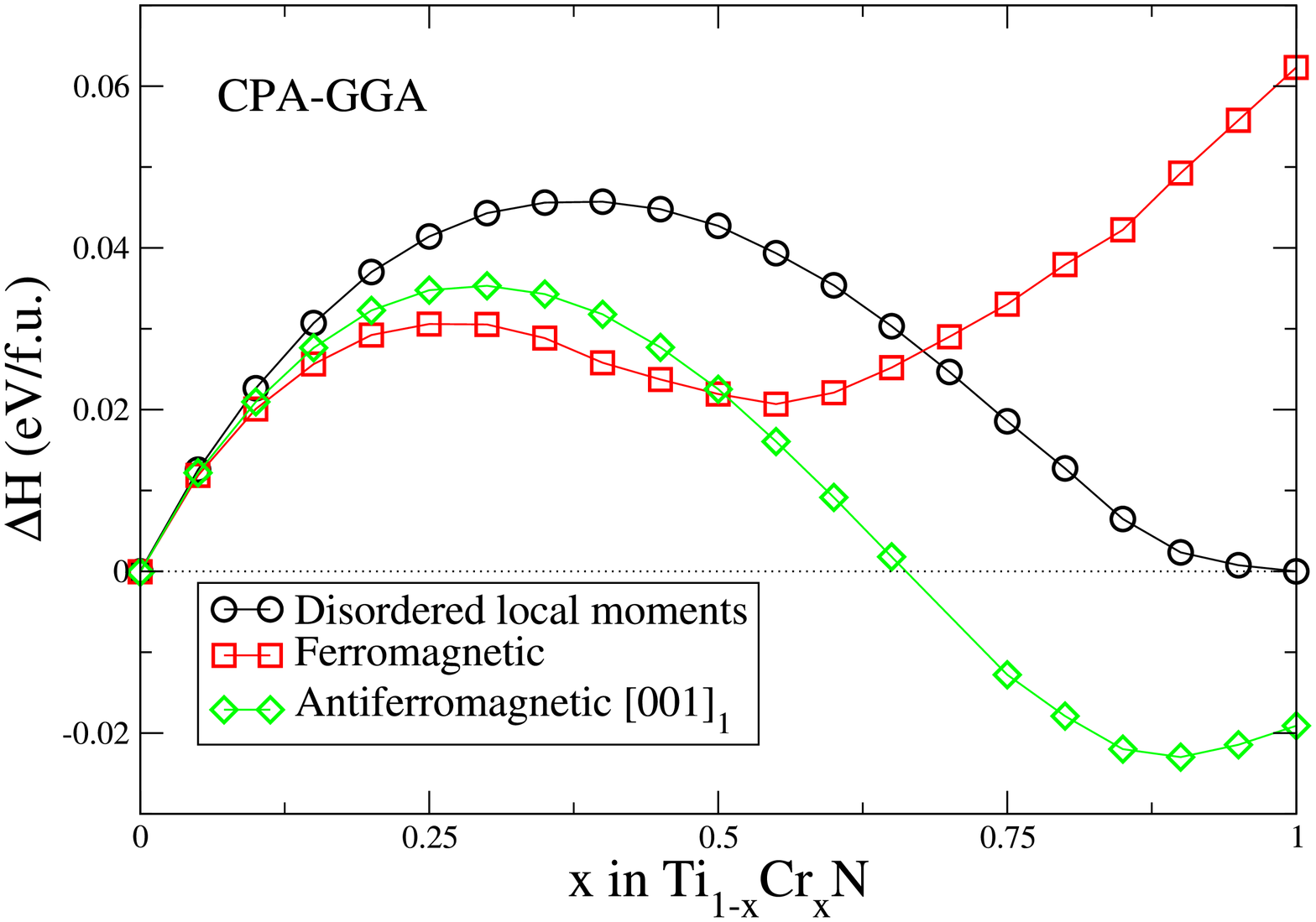}
\caption{\label{fig:CPA_H} (Color online) The calculated mixing enthalpies of cubic ferromagnetic (squares), cubic antiferromagnetic [001]$_1$ (diamonds), and cubic disordered magnetic (circles) Ti$_{1-x}$Cr$_{x}$N as calculated with the KKR-ASA method and the CPA treatment of disorder using the GGA exchange-correlation functional. The lines are guides for the eye.}
\end{figure}

To get a more detailed picture of the relative energies of the cubic phases and to be able to simulate the disordered magnetic state over the whole composition range, we have calculated the mixing enthalpies of Ti$_{1-x}$Cr$_x$N using the KKR-ASA method employing the CPA approximation to model the solid solution. The results are presented in Fig.~\ref{fig:CPA_H}. The values of mixing enthalpies are in good agreement with the SQS calculations using GGA in Fig.~\ref{fig:SQS_H} but provides the full picture over a dense concentration mesh including the relation between ordered and disordered magnetic states. Despite the overall qualitative agreement, there are two differences that can be noted: Firstly the FM state for pure CrN is not as high in energy relative both the disordered magnetic and the AFM state in the KKR-ASA calculations as compared to the PAW calculations. This could be due to the larger approximations done in the ASA for the one-electron potential. However, we note that for $x\leq 0.75$ the enthalpies are very close between the two methods and the enthalpies of the AFM[001]$_1$ state is similar for the entire composition range. Secondly, the CPA calculations neglects the effect of local lattice relaxations which has a maximum value of about $0.020$~eV/f.u. at the composition $x=0.50$. Since the local lattice relaxation energies are very similar in the FM and AFM[001]$_1$ state, this neglect in the CPA calculations do not affect our conclusions about magnetism. 

In the CPA calculation the FM state becomes lower in energy than the AFM[001]$_1$ state more or less at $x=0.50$. For lower $x$ the FM state is lowest in energy but, perhaps surprising, also the AFM[001]$_1$ state is considerably below the disordered magnetic state in energy. At very high TiN content, as the Cr atoms becomes very diluted, the magnetic states becomes almost degenerate in energy. This is the case also if we consider the energy per Cr atom rather then per formula unit as it is shown in the figure.

Our calculations thus support the experimental reports of ferromagnetism in TiN-rich compositions of Ti$_{1-x}$Cr$_x$N solid solutions. The exact transition point between the FM and AFM[001]$_1$ states are slightly dependent on the details of our modelling: Almost at $x=0.50$ according to the GGA calculations, or around $x=0.65$ in the LDA+U calculations. However, in the present work we have not considered local environment induced spin-flip states, shown to be of importance for the understanding of the Invar effect in FeNi alloys~\cite{Abrikosov2007, Liot2009}. Such states, where certain Cr-moments in Cr-rich environments are anti-parallel to the net magnetization, could very well be the actual groundstate for a composition range around the FM to AFM transition.  

With the mixing enthalpies at hand, we can consider whether or not the chemical solid solution reported in the experiments are likely to be the thermodynamic equilibrium or if there is a tendency  towards chemical decomposition or order. If we make a simple mean field analysis adding the entropy of the ideal solid solution to the DLM mixing enthalpy of Fig.~\ref{fig:CPA_H}, we find that a temperature of about 1400 K would be needed to fully stabilize the solid solution with respect to phase separation over the whole concentration range. Since the CPA calculations neglect local lattice relaxations and that the mean field analysis typically overestimates transition temperatures it is likely that the bulk samples in Ref.s.~\cite{Aivazov1975, Inumaru2008} that are synthesized at 1373 and 1273~K respectively could really be solid solutions. However, one can not exclude the presence of non-negligible short range order or clustering tendencies on the metal sublattice, especially around the composition $x=0.40$. Even if the positive mixing enthalpies indicate overall clustering tendencies, there could still be correlation shells that show ordering tendencies. The local lattice relaxation of N atoms, taking place significantly only in between next-nearest metal neighbors of different kinds was discussed in Ref.~\cite{Alling2007} to give rise to such effects on the second metal coordination shell of Ti$_{1-x}$Al$_x$N and could be of importance also in the present case.

 In the solid solution promoting thin film synthesis in Ref.~\cite{Inumaru2007}, atomic diffusion is probably limited during growth due to the rather low substrate temperature (973~K) but some degree of short range clustering or ordering tendencies could be present also in those solid solutions.

\subsection{Exchange interactions and the magnetic phase diagram}
\begin{figure}
\includegraphics[angle=-90,width=0.92\columnwidth]{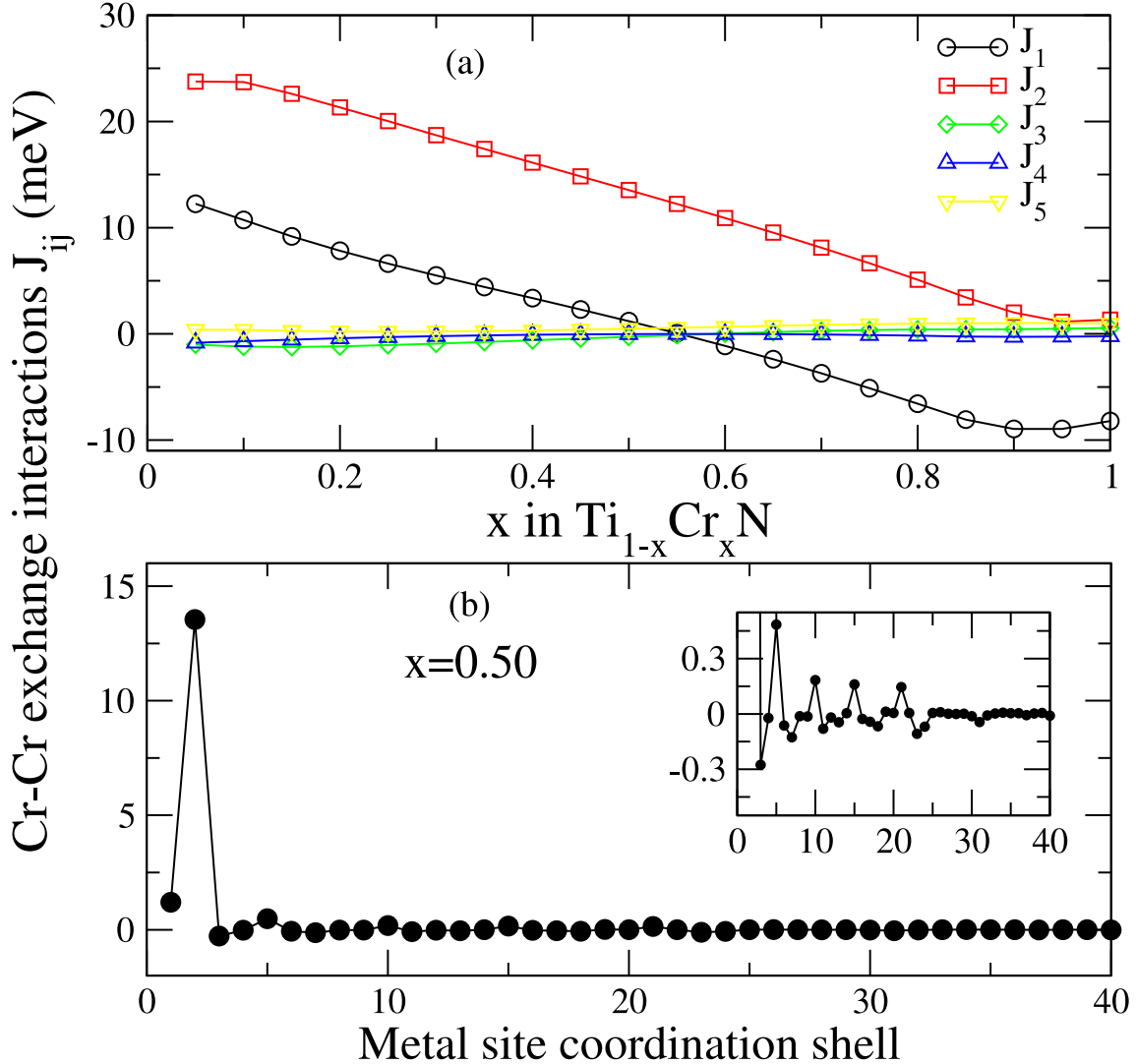}
\caption{\label{fig:Jij} (Color online) (a) The calculated Cr-Cr magnetic exchange interactions, $J_{ij}$ for the first 5 coordination shells as a function of the composition. (b) The 40 first interactions for the fixed composition $x=0.50$. The inset shows a zoom in of the weaker interactions.}
\end{figure}

To obtained increased understanding of the physics that changes the energetic preferences of different magnetic states with composition we calculate the magnetic pair exchange interactions $J_{ij}$ of the Heisenberg hamiltonian in Eq. 1 of the cubic lattice as a function of composition. The first 8 interactions are calculated for all compositions and the first 40 interactions have been obtained for $x=0.50$ and $x=0.30$. The results for the first 5 coordination shells are presented in panel (a) of Fig.~\ref{fig:Jij}. The system is strikingly dominated by the first two interactions denoted $J_1$ and $J_2$. In pure CrN we obtain $J_1= -8.2$ meV while $J_2 = 1.3$ meV. This can be compared to the values obtained in Ref.~\cite{Filippetti2000}, $J_1 =-9.5$ meV and $J_2=4$ meV with a structure inversion method where all other interactions were neglected. Of course the strong negative interaction on the first shell guarantees an antiferromagnetic ground state in the case of pure CrN on the cubic lattice without orthorhombic distortions as was discussed by Filippetti \emph{et al.}~\cite{Filippetti1999, Filippetti2000}.

 When the amount of TiN is increased (decreasing $x$) in the system, the first two exchange interactions are strongly influenced. Both increase almost linearly with increasing TiN content. $J_1$ changes sign close to $x=0.55$ and becomes positive for higher TiN content. A maximum value of $J_1=12.2$ meV is obtained at $x=0.05$. $J_2$ increases in a similar manner and reaches values of 13.5 meV at $x=0.5$ and 23.8 meV at $x=0.05$. The more long ranged interactions also changes slightly but they all stay close to 0 for all compositions. 
 
 Here it is important to note that the strongest interaction, $J_2$, does not influence the energy difference between the ferromagnetic and the AFM[001]$_1$ configuration. Both these two spin configurations show parallel alignment of all spins on the second coordination shell of the metal fcc-sublattice. So even though a large positive value of $J_2$ favors ferromagnetism with respect to many other configurations, including the completely disordered state shown in Fig.~\ref{fig:CPA_H}, it is the other interactions that decides which of FM and AFM[001]$_1$ that is the magnetic ground state. In particular the change of sign of $J_1$ is important to decide the transition between FM and AFM[001]$_1$, although close to this point when $J_1$ is very small, also weaker long range interactions have a quantitative influence.
 
 Panel (b) of Fig.~\ref{fig:Jij} shows the $J_{ij}$ on the first 40 coordination shells for the composition $x=0.50$. In this case the second coordination shell interaction is one order of magnitude larger than the interaction of the first coordination shell and almost 2 orders of magnitude larger as compared to the more long ranged and oscillating interaction of which a zoom-in is shown as an inset in the figure.
 
Before we go into a discussion about the electronic origin of these changes, we study their thermodynamic consequences by simulating the magnetic phase diagram using Monte Carlo simulations to derive the magnetic critical temperature for all compositions on the cubic lattice.

\begin{figure}
\includegraphics[angle=-90,width=0.87\columnwidth]{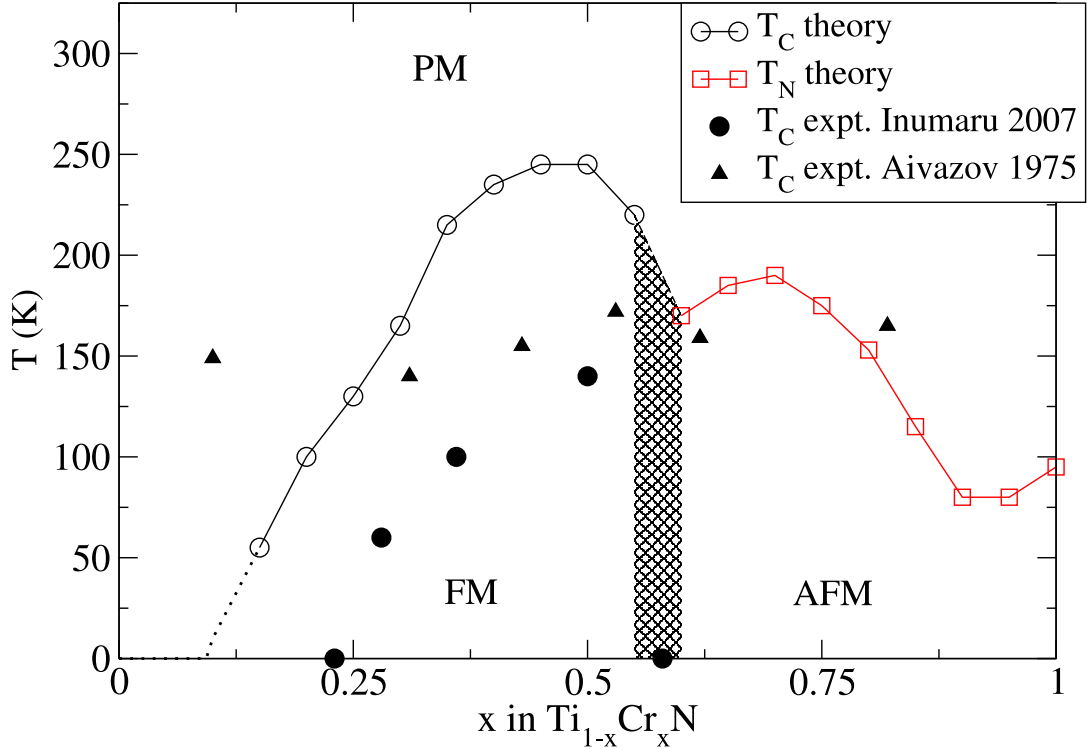}
\caption{\label{fig:Tc} (Color online) The calculated magnetic phase diagram of cubic Ti$_{1-x}$Cr$_x$N solid solutions. The calculated temperatures of the ferro-to-paramagnetic transition ($T_C$) are shown with open circles. The calculated temperatures of antiferro-to-paramagnetic transitions ($T_N$) are shown with open squares. The experimental values of $T_C$ from Inumaru \emph{et al.}\cite{Inumaru2007}~are shown with solid circles while the values from Aivazov and Gurov~\cite{Aivazov1975} are shown with solid triangles.}
\end{figure}

Figure~\ref{fig:Tc} shows the calculated magnetic phase diagram of cubic B1 Ti$_{1-x}$Cr$_x$N solid solutions. The calculated  ferromagnetic to paramagnetic transition temperatures, the Curie temperature (T$_C$), are shown with open circles while the antiferromagnetic to paramagnetic transition temperature, the N\'eel temperature (T$_N$)  are marked with open squares. The transition region from FM to AFM ordering according to these calculations  takes place between x=0.55 and 0.60, this is marked with a shaded region. The experimental values reported by Aivazov and Gurov~\cite{Aivazov1975} and Inumaru~\emph{et al.}~\cite{Inumaru2007} are shown with solid triangles and circles respectively.

At low Cr content, $x < 0.15$, despite the relatively strong exchange interactions, no magnetic ordering can be deduced from the calculations. Of course, this is due to the strong dilution of Cr atoms in the lattice. At a Cr content of $x=0.15$ we obtain $T_C=55~K$ and $T_C$ then rapidly increase with $x$ reaching a maximum value of $T_C = 245~K$ at $x=0.45$ and $x=0.50$. At this point the decrease in interaction strengths overcomes the effect of gradual increase in Cr content and $T_C$ decreases to $T_C= 220~K$ at $x=0.55$. At this composition, $J_1$ changes sign and  we start to obtain AFM ordering from $x\geq 0.60$. In the AFM region $T_N$ first increase with increasing Cr content reflecting the larger negative values of $J_1$. The maximum $T_N=190~K$ on the cubic lattice is reached at $x=0.70$ and at even higher Cr content the critical temperature decrease with $x$. The critical temperature on the cubic lattice of pure CrN is $T_N = 95~K$. Of course, in experiments the orthorhombic distortion takes place in Cr-rich samples and the theoretical values of $T_N$ on the fixed cubic lattice for those cases are included only for completeness. 

When comparing our theoretical results with the experimental measurements of Inumaru~\emph{et al.}~\cite{Inumaru2007} (solid circles in Fig.~\ref{fig:Tc}) a qualitative agreement can be seen in the Ti-rich region but two important differences needs to be discussed.

First, the theoretical values of $T_C$ with $x \leq 0.50$ overestimate the experimental critical temperatures with about 100 $K$. Indeed many effects contribute to the difficulty to derive the magnetic critical temperature from first principles with quantitative accuracy as we discussed in details in Ref.~\cite{Alling2009}. Thermal expansion and other vibrational related phenomena, as well as electronic excitations and structural defects are of importance~\cite{Alling2009}. Of course also inaccuracies in the approximations used for electronic exchange-correlation effects can have an influence. In addition to those effects, we believe that there could be a quantitative impact of local environment effects on the magnetic interactions between Cr atoms even if there is none or only a moderate level of short range clustering or ordering present. In the solid solution there exist Cr atoms in Ti-rich and Cr-rich environments. When we derive the magnetic interactions within the CPA framework we miss the fact that in Cr-rich environments, the interactions should tend to those we obtain at Cr-richer global compositions. In the same way, in Ti-rich environments, the interactions should tend in the directions of the values we obtain for Ti-richer global compositions. In the Monte Carlo simulation, we treat all interactions between Cr-atoms as equal to the mean value obtained from the CPA calculations. Cr rich environments will thus develop magnetic short range order at somewhat higher temperatures in our simulations as compared to the real experimental situation. Even though this is somewhat counteracted by the Cr-poor environments that are treated with slightly to weak interactions in our simulations, the net effect is likely to be an overestimation of $T_C$. Another possible explanation is that there are short range chemical ordering tendencies on the second coordination shell, possibly induced by the N-atom local relaxation effect discussed in Ref.~\cite{Alling2007}. If that is the case, Cr-atoms would be less likely to have other Cr atoms present on the coordination shell where the magnetic interactions are the strongest. That would lead to a lower $T_C$ as compared to the ideal solid solution case considered in the present simulations. 

The second difference is that Inumaru~\emph{et al.} obtain no magnetic ordering at $x=0.58$ and visualize this with writing $T_C=0~K$~\cite{Inumaru2007}. At first glance this seems at odds with our results, but this is not the case. Inumaru~\emph{et al.} derived the critical temperature from the temperature of onset of net magnetization in the presence of a magnetic field. Their sample is simply likely to be antiferromagnetic thus showing no net global magnetization. At this point it is suitable to note that in our case, in contrast to the single parameter Heisenberg Hamiltonian or a case where the first interaction of the fcc sublattice dominates, the critical temperature of the phase transition would not go to zero between the FM and AFM regions. The large value of $J_2$ ensures that we would se a peak in $C_V$ at finite temperature also in the shaded area of Fig.~\ref{fig:Tc}.  

When instead comparing with the experimental values obtained by Aivazov and Gurov~\cite{Aivazov1975} the difference is more striking since they observe ferromagnetic transitions with $T_C\geq 140~K$~over a concentration interval of $0.10 \leq x \leq 0.82$. The most likely explanation of this discrepancy with both our calculations and the later measurements by Inumaru~\emph{et al.} is that they obtained inhomogenous samples with local domains with varying compositions large enough to give independent signals in the susceptibility measurements.  

\subsection{Electronic structure}

Our investigation has this far confirmed the transition from antiferromagnetism in CrN to ferromagnetism in TiN rich Ti$_{1-x}$Cr$_{x}$N solid solutions, revealed the changes of magnetic exchange interaction with composition and illuminated some interesting aspects of the magnetic thermodynamics in the system. In order to understand these changes on the most fundamental level of physics we turn to a study of the electronic structure of Ti$_{1-x}$Cr$_x$N.

\begin{figure}
\includegraphics[angle=-90,width=0.87\columnwidth]{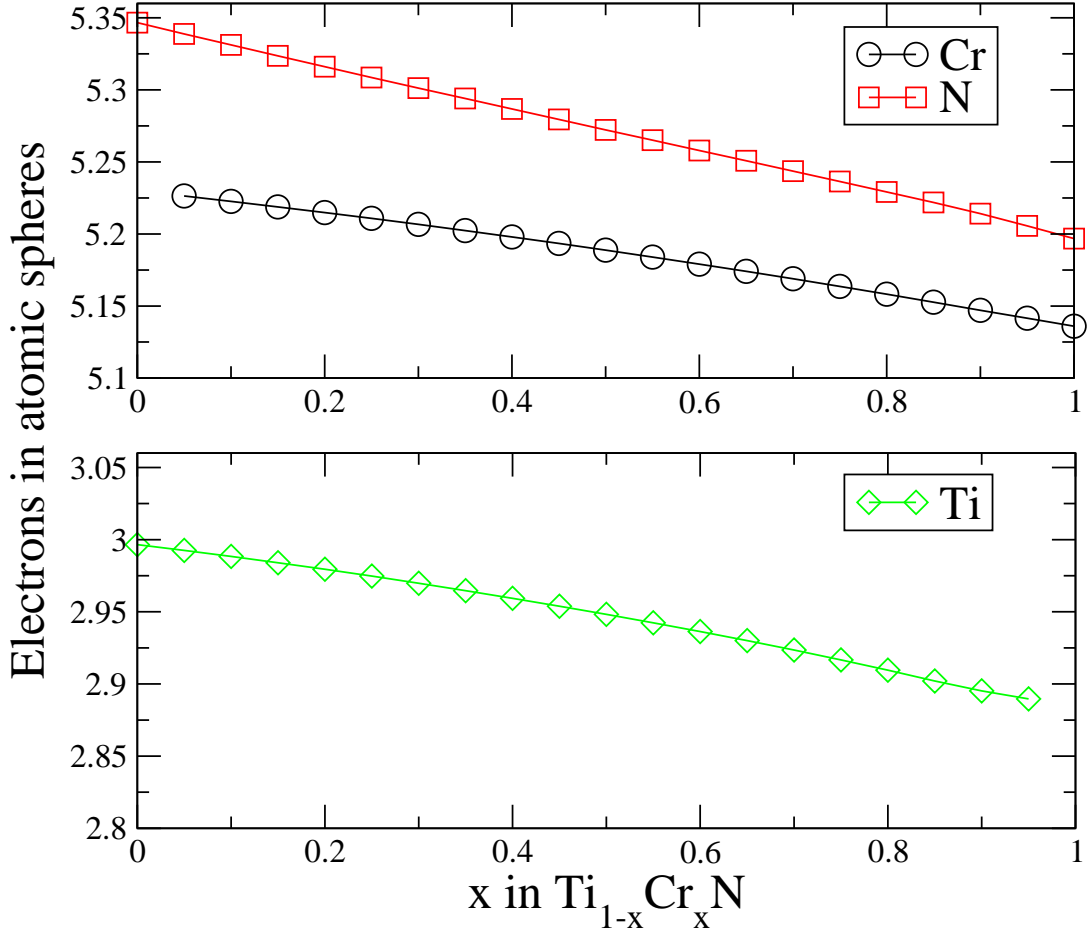}
\caption{\label{fig:Q} (Color online) The amount of electron density inside the N, Cr, and Ti atomic spheres in the KKR-ASA-CPA calculations. }
\end{figure}

Before we enter into a discussion about the particular electron density of states and their importance for magnetism, we study the redistribution of electronic charge in the system as a function of composition. Fig.~\ref{fig:Q} shows the electron density inside the Ti, Cr, and N atomic spheres in our KKR-ASA calculations. The figure shows that Ti in TiN loses more electron density to the nitrogen as compared to Cr in CrN. It also shows that in the alloy, there is a net transfer of charge from Ti to Cr. The higher the TiN concentration in the system, when each Cr atom has more Ti metal neighbors, the more extra charge is accumulated around the Cr nuclei. This fact, that in the Ti$_{1-x}$Cr$_x$N solid solutions, the magnetic interactions between Cr-moments take place in an electron rich environment is an important clue to the origin of the changes observed in the previous sections. The total electron density is balanced to 9 electrons in pure TiN and 11 electrons in CrN through an almost constant electron density in the interstitial region, represented in the calculation by two atomic spheres without nuclei. 

\begin{figure}
\includegraphics[angle=-90,width=0.87\columnwidth]{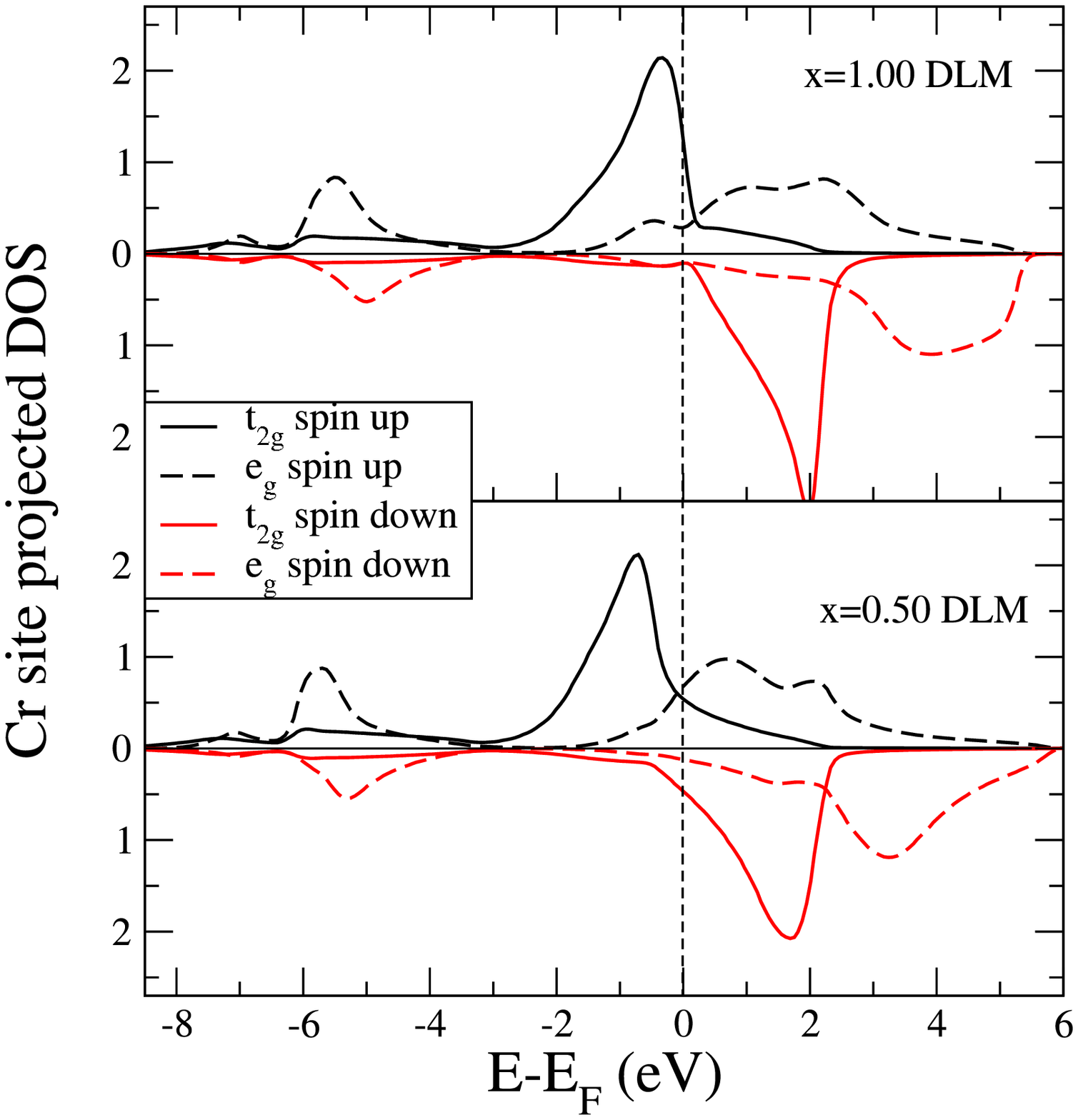}
\caption{\label{fig:CPA_DOS} (Color online) The Cr-site projected electronic density of states with 3d character in the DLM magnetic state of pure CrN (top panel) and Ti$_{0.5}$Cr$_{0.5}$N (lower panel) as calculated with the KKR-ASA-CPA method using the GGA exchange-correlation functional. $t_2g$-states are shown with solid lines and $e_g$-states with dashed lines.}
\end{figure}

The electronic density of states of CrN have been studied by Filippetti~\emph{et al.}~using LDA~\cite{Filippetti1999} and more recently by Herwadkar~\emph{et al.}~using the LDA+U method~\cite{Herwadkar2009}. The main features are similar to other B1 transition metal nitrides and carbides like TiC, TiN, and ScN~\cite{Price1989, Alling2008Surf} where a split of the 3d-states according to their symmetry takes place with $e_g$-states forming strong covalent bonds with nearest neighbor N (or C) 2p-states giving rise to bonding-anti-bonding bands, while $t_{2g}$-states that are oriented away from the N (or C) atoms form a more narrow non-bonding band roughly in between. The occupation of the non-bonding band depends on the valence of the system ranging from empty in ScN to half-filled in CrN~\cite{Alling2008Surf}. Due to magnetism in the latter case, this band is split and the spin up part is almost or completely filled depending on the exchange correlation functional~\cite{Herwadkar2009}.  Correspondingly the $e_g$-character anti-bonding band is almost or completely empty. 

The top panel in Fig.~\ref{fig:CPA_DOS} shows the Cr site and symmetry projected electronic density of states (DOS) with d-character in pure CrN ($x=1.00$) in the DLM magnetic state using the KKR-ASA method and GGA functional. The splitting between $e_{g}$ and $t_{2g}$ states is clearly seen as well as the magnetic impact. The spin up non-bonding state is almost entirely filled while the spin up anti-bonding state has a marginal occupation in line with previous GGA and LDA calculations~\cite{Filippetti1999, Alling2008Surf}. It should be noted that the slight overlap between the spin orientations of the $t_{2g}$ state is due to overlapping tales from the states of neighboring atoms with opposite spin orientation. The lower panel of Fig.~\ref{fig:CPA_DOS} show the situation in the Ti$_{0.50}$Cr$_{0.50}$N solid solution. The excess of electron density on the Cr sites can not be understood as only tails from neighboring Ti-states. Instead there is a more or less rigid band shift with respect to the Fermi level due to a larger occupation of spin up $e_g$ anti-bonding states as well as spin down $t_{2g}$ non-bonding states.

\begin{figure}
\includegraphics[angle=-90,width=0.87\columnwidth]{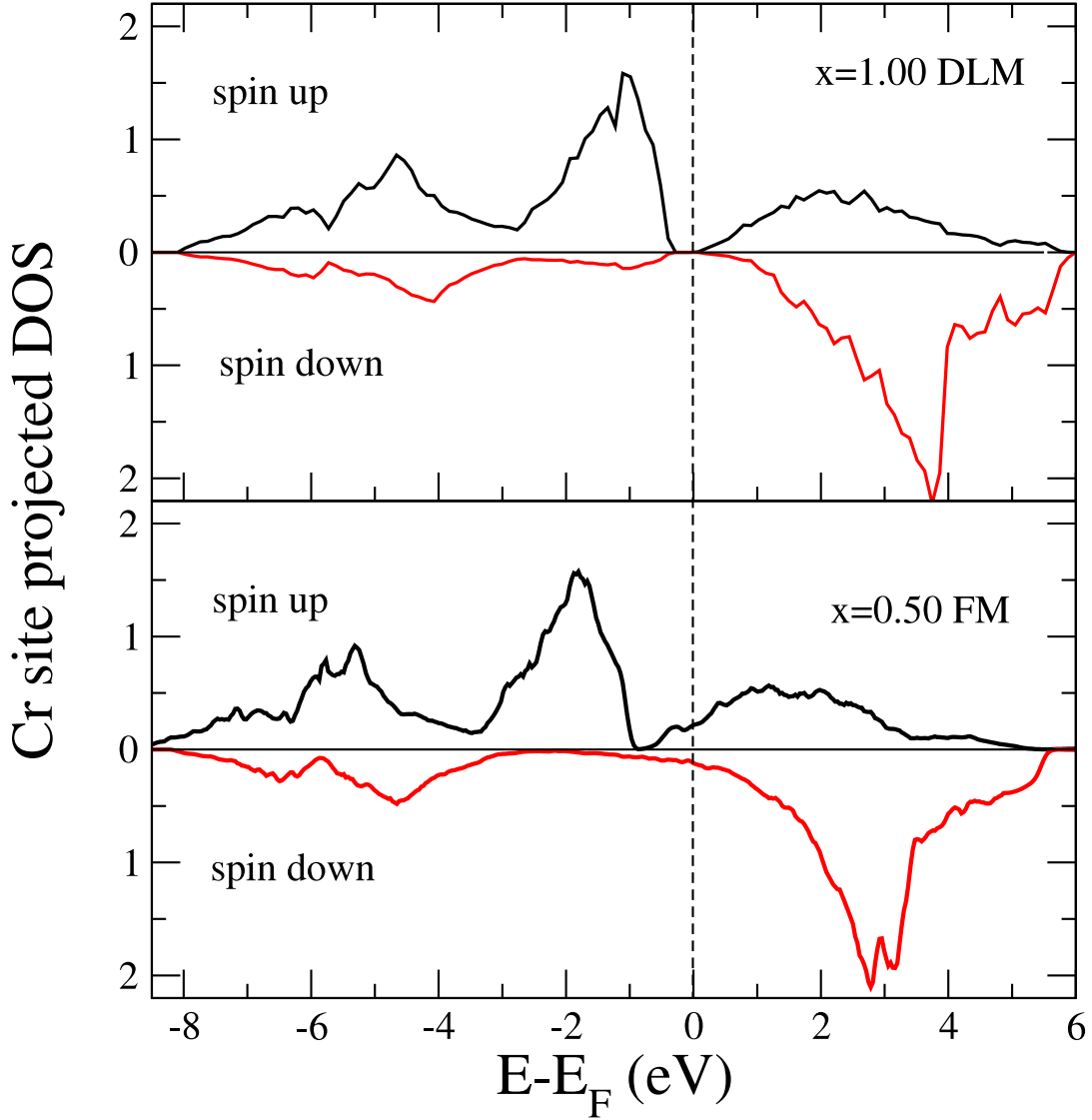}
\caption{\label{fig:VASP_DOS} (Color online) The Cr-site projected electronic density of states of the DLM magnetic state of pure CrN (top panel) and in the FM magnetic state of Ti$_{0.5}$Cr$_{0.5}$N (lower panel) calculated with the PAW-SQS method and the LDA+U treatment of exchange and correlation effects.}
\end{figure}

When taking strong electron correlation into account on the level of the LDA+U approximation the story becomes even more clear. Fig~\ref{fig:VASP_DOS} shows the Cr site projected total electronic density of states as calculated with the PAW-SQS method and the LDA+U approach to exchange-correlation energies. The top panel shows the DOS for pure CrN in the disordered magnetic state resembling the KKR-ASA-GGA calculations discussed above, but with a more distinct separation between the non-bonding spin up on one hand and anti-bonding spin up as well as non-bonding spin down states on the other, with a small bandgap in between. The lower panel shows the Cr-site projected total electron DOS of Ti$_{0.50}$Cr$_{0.50}$N in the FM state displaying the same band shift and the onset of occupation of the spin up anti-bonding state and spin down non-bonding state as seen above. These results suggest that even if there is a slight impact of the difference between the GGA and LDA+U treatment of exchange-correlation effects in pure CrN, the difference should be smaller at higher TiN content when the features around the small bandgap is pushed down below $E_F$. 

The importance of these changes in band filling for the magnetic interactions can be understood in the following way: The increased partial occupation of the $e_g$ anti-bonding state promotes the double-exchange mechanism that favors ferromagnetic coupling since such an arrangement decrease the kinetic energy of the $e_g$ electrons through delocalisation over the Cr-N-Cr bond, possibly affecting not only next-nearest neighbor interactions but also nearest neighbors. The increased electron density on the N-sites could also promote this effect. The increased occupation of the spin down non-bonding state should counteract the antiferromagnetic interaction due to overlap of the $t_{2g}$ states of Cr-Cr neighbors. Taken together these effects qualitatively explains the change in magnetic interactions observed in Fig.~\ref{fig:Jij}. However, as always in real materials, other magnetic effects are also present like the oscillating RKKY-type interactions of Cr moments mediated through conduction electrons of TiN origin, as seen in the inset in the lower panel of Fig.~\ref{fig:Jij}.

\section{Conclusions}
In conclusion we have studied the magnetism of the Ti$_{1-x}$Cr$_x$N solid solutions by means of first-principles calculations and confirmed the experimental findings of ferromagnetic ordering in the TiN-rich regime corresponding to approximately $x \leq 0.50$. The orthorhombic distortion associated with the  [011]$_2$~antiferromagnetic state is only favorable in the most CrN-rich regime while a [001]$_1$ antiferromagnetic state with negligible tetragonal distortions of the B1 lattice is the most favorable magnetic ordering in an intermediate region. The Cr-Cr magnetic interactions on the first two coordination shells are heavily influenced by TiN addition and increase linearly with increasing TiN composition. However the Curie temperature anyway decrease for $x\leq 0.40$ due to the dilution of Cr atoms in the lattice. The change in magnetic interactions originate in a charge redistribution from Ti to Cr and N in the solid solutions. It leads to a strengthening of the ferromagnetic double-exchange mechanism as the anti-bonding Cr 3d states with $e_g$-symmetry are increasingly populated with increasing TiN content. At the same time the antiferromagnetic interactions originating from the overlap of Cr $t_{2g}$-states are weakened as their spin polarization decrease with increasing TiN content. The understanding of these effects could be used as a guide for future attempts to design hard nitride materials that are ferromagnetic at room temperature.

\section{Acknowledgement}
The Swedish Research Council (VR) is acknowledged for financial support.
Most of the calculations were performed using computational resources allocated by the Swedish National Infrastructure for Computing (SNIC).





%

\end{document}